\documentclass[preprint,12pt]{aastex}
\usepackage{epsfig}
\usepackage{emulateapj5}
\usepackage{apjfonts}

\begin{document}
\submitted{Submitted to ApJ, 22 April 2005}
\def\num1{${\cal D}_1$}
\def\mess{${{\theta_E}\over{\theta_1}}$}
\def\tetae{{$\tau_1 ({\rm days})$}}  
\def\br1br{${\cal R}_1^b ({\rm yr}^{-1})$}  
\def\r1d{${\cal R}_1^d ({\rm yr}^{-1})$}  
\def\rtb{{\bf ${\cal R}_{tot}^b ({\rm yr}^{-1})$}}  
\def\rtd{\bf{${\cal R}_{tot}^d ({\rm yr}^{-1})$}}   
\def\gc{globular cluster}
\def\kms{{\rm km/s}}
\def\rmeq#1{\eqno({\rm #1})}
\def\gl{gravitational lens} \def\gb{Galactic bulge}
\def\lc{light curve}
\def\ml{microlensing} \def\mo{monitor}
\def\pr{program} \def\mlmpr{\ml\ \mo\ \pr}
\def\ev{event}
\def\ex{expansion}
\def\fn{function} \def\ch{characteristic}
\def\bi{binary}  \def\bis{binaries}
\def\rd{Di\thinspace~Stefano}
\def\bl{binary lens} \def\dn{distribution}
\def\pop{population}
\def\ct{coefficient}
\def\cc{caustic crossing}
\def\cs{caustic structure}
\def\mag{magnification}
\def\pl{point lens}
\def\dm{dark matter}
\def\rd{Di\thinspace~Stefano}
\def\tots{track of the source}
\def\de{detection efficiency}
\def\det{detection}
\def\ob{observ}
\def\ol{observational}
\def\od{optical depth}
\def\ml{microlensing}
\def\mtm{monitoring team}
\def\mmtm{microlensing monitoring team}
\def\otm{observing team}
\def\mo{monitor}
\def\motm{microlensing observing team}
\def\los{line of sight}
\def\ev{event}
\def\by{binarity}
\def\ptb{perturb}
\def\sgf{significant}
\def\bis{binaries}
\def\sg{signature}
\def\bbl{binary-lens}
\def\kms{{\rm km/s}}
\def\rmeq#1{\eqno({\rm #1})}
\def\gl{gravitational lens} \def\gb{Galactic bulge}
\def\lc{light curve}
\def\ml{microlensing} \def\mo{monitor}
\def\pr{program} \def\mlmpr{\ml\ \mo\ \pr}
\def\ev{event}
\def\ex{expansion}
\def\fn{function} \def\ch{characteristic}
\def\bi{binary}  \def\bis{binaries}
\def\rd{Di\thinspace~Stefano}
\def\bl{binary lens} \def\dn{distribution}
\def\pop{population}
\def\ct{coefficient}
\def\cc{caustic crossing}
\def\cs{caustic structure}
\def\mag{magnification}
\def\ppl{point-lens}
\def\bl{blending}
\def\mage{magnification}
\def\fsse{finite-source-size-effects}
\def\cc{caustic crossing}
\def\mp{multiple peak}
\def\lcf{lightcone fluctuation}
\def\wdf{white dwarf}
\def\pn{planetary nebula}
\vskip -.8 true in
\title{
Mesolensing Explorations of Nearby Masses:
From Planets to Black Holes}

\author{Rosanne Di\thinspace~Stefano}
\affil{Harvard-Smithsonian Center for Astrophysics, 60
Garden Street, Cambridge, MA 02138} 
\affil{Kavli Institute for Theoretical Physics, University
of California Santa Barbara, Santa Barbara, CA 93106} 

\def\gl{gravitational lensing}
\def\Gl{Gravitational lensing}
\def\ml{microlensing} 
\def\Ml{Microlensing} 
\def\Et{Einstein angle} 
\def\et{$\theta_E$} 
\def\Er{Einstein radius}
\def\ev{event}
\def\vb{variable}
\def\vy{variability}
\def\sg{signature} 
\def\asec{arcsecond}

\begin{abstract}
Nearby masses can have a high probability of lensing stars
in a distant background field. High-probability lensing, or mesolensing,
can therefore 
be used to dramatically increase our knowledge of
dark and dim objects in the solar neighborhood, where it can 
discover 
and study members of the local dark population (free-floating
planets, low-mass dwarfs,
white
dwarfs, neutron
stars, and stellar mass black holes).
We can measure the mass and transverse velocity of
those objects discovered (or already known), 
and determine whether or not they are in binaries with dim companions.  
We explore these and other applications of mesolensing, including
the study of forms of matter that have been hypothesized but not
discovered, such as
intermediate-mass black holes,
dark matter objects free-streaming through the Galactic disk,
and planets in the outermost regions of the solar system. 
In each case we discuss the feasibility of deriving results
based on present-day monitoring systems, and also consider the 
vistas that will open with the advent of all-sky monitoring in the 
era of
the {\sl Panoramic Survey Telescope and Rapid Response System} (Pan-STARRS)
and the {\sl Large Synoptic Survey Telescope} (LSST).    
\end{abstract} 
\submitted{Submitted to ApJ, 22 April 2005}

\section{Detecting Mesolensing}

The companion paper introduces the idea of mesolensing, in which
a single lens has a high probability of producing a detectable event
(Di\, Stefano 2006a).
Nearby stellar remnants are examples of mesolenses. 
Consider, e.g., a 
neutron star located $100$~pc from us, traveling with
a transverse speed of $100$~km~s$^{-1}$ in front of
M31. Over the course of
twenty to thirty years, 
the Einstein ring of the neutron star will traverse an area
of roughly $0.02'' \times (4-6)''.$ In most regions of M31, this area
will contain several hundred stars. It is likely that one or more of these
stars is bright enough that, when it is magnified, 
the
integrated luminosity from a square arcsecond region around it will
increase by $10\%$.
During the same interval, there may also
be several smaller magnifications,
producing jitter in the baseline light. In 
addition, astrometric shifts of
$\sim 0.001''$ may occur and be detectable in the images of 
several stars lying in an even larger swath across the sky.   
 These lensing effects could potentially
lead, through a monitoring program,
to the discovery of 
a ``new" neutron star, one whose presence is not yet known. 
Alternatively, if there is a known pulsar or 
X-ray emitter along the direction
to M31, the region behind the mass can be
targeted for repeated deep, high-resolution observations.
Such observations would have the potential to 
measure the mass of the neutron star, determine its proper motion,
and probe for evidence of companions.  

This type of argument can be applied to other dark and dim masses
located within roughly a kpc, 
ranging from the relatively rare isolated black hole, to the more ubiquitous
low-mass dwarfs. If the lensing action of such objects can be detected,
gravitational lensing will become a valuable tool to discover a larger 
fraction of them and to study their properties, 
including their masses and multiplicities.  

\subsection{Plan of the Paper} 

The primary goal of this paper is to 
study applications of mesolensing. We focus exclusively
on photometric effects, although astrometric effects are also expected.
The necessary background
is covered in \S 2.  
Section 3 focuses on computing the event rates due to nearby
stars: low-mass dwarfs, white dwarfs (WDs), neutron stars (NSs), and
black holes (BHs). These forms of matter are known to exist.
The predicted rates are high enough to provide a guaranteed signal in
past, ongoing, and future monitoring programs.
This signal can be used to study these nearby populations
in a way that has not so far been possible. 
Even if the primary goal of a monitoring program
 is to study dark matter, 
the signal due to nearby
masses must be identified   
so that events due to nearby lenses
can be subtracted from the total lensing
signal to determine the true contribution of other lens
populations, such as MAssive Compact Halo Objects (MACHOs).  
 
In \S 4, attention turns to forms of nearby matter for which
there is as yet no direct evidence. Specifically, 
we consider (1)~intermediate-mass black holes (IMBHs),
(2)~disk dark matter in the form of free-streaming planet-mass
objects, and (3)~possible planets in the outer regions 
($> 1000$ AU) of the solar system.  
Section 5  is a summary of the conclusions.     
  
\section{Background on Mesolensing}

Lensing events have been discovered by
monitoring programs, each of which regularly observes a 
dense background field 
at intervals of one to several days. 
Attention so far has focused on
 the Galactic Bulge and the 
Magellanic Clouds. M31 has also been subjected to regular
monitoring. 
The monitoring programs are designed to discover evidence of lensing
by masses lying in front of the background; the presence of the specific
masses that eventually produce events is not generally known prior
to the observations. Monitoring observations can therefore discover and
probe the properties of a population of dark or dim objects. 

The monitored  
field can be viewed as a composite of many small subfields,
each with linear dimensions $\theta_{mon},$ where the value of
$\theta_{mon}$ has typically been on the order of an arcsecond.
The term ``monitored region", for a specific lens, refers to a box
centered on the lens with area   
 $\theta_{mon} \times \theta_{mon}.$ It was necessary to develop
a mode of analysis that considers  
monitored regions, rather than individual stars,
because the dense fields we observe always have 
multiple stars within each resolution element 
(Di\thinspace Stefano \& Esin 1995).   
Although the details of the analysis
can be complex, the basic idea is that  
the image of the region obtained at time $t$ is compared with
an appropriately averaged image of the field, and differences
are identified. The value of this difference imaging approach is that 
variations of an individual star can be discovered, even when the field
is so dense that the
lensed star itself cannot be imaged by the telescope used
for monitoring 
(see, e.g., Alard \& Lupton 1998).

\subsection{Detectable Photometric Events}

\noindent{\bf Fractional increase in the amount of light received:} In 
order for a photometric deviation above baseline to be
detectable, we must receive an additional amount of light per unit
time from the region, producing a fractional change of at least $f_T.$
Typical values of $f_T$ used so far in monitoring programs have been
in the range $0.1-0.5.$

\noindent{\bf Distance of closest approach:} Let 
the index $i$ label
an individual source star that lies close to the path of the lens.  
If the approach between this source and the lens is to produce a 
detectable photometric event, 
the required angular
distance  of closest approach is 
$\theta_{b,i}.$ 
Because it is 
convenient to express
angles in terms of the Einstein angle of the lens, we define
$b_i = \theta_{b,i}/\theta_E.$  
If an individual star must experience a magnification $>>1$ in order
for the magnification of the monitored region to be $(1+f_T)$, then
\begin{equation}
{b_i} = {{L_i}\over{\langle L \rangle}}\, {{1}\over{f_T N_{mon}}}
\end{equation}
This limit of small $b_i$ is the limit used to compute the event rates
presented below. It is possible, however for $b_i$ to be even larger than
unity. If, e.g., a single bright star is the dominant source
of light in the monitored region, then $b_1 = 1, 2.8, 3.5$ 
corresponds to $f_T$ approximately equal to $0.34, 0.02, 0.01,$
respectively. 
  
\noindent{\bf Event Duration:} For 
point-source/point-lens events,
the event duration is the only measurable parameter related to the lens mass.   
The Einstein crossing time is 
\begin{eqnarray}
\tau_E &=&{{2\, \theta_E}\over{\omega}} \nonumber\\
       &=&35{\rm days}\, \Big({{M}\over{1.4\,  M_\odot}}\Big)^{{1}\over{2}}
          \Big({{100\, {\rm km\, s^{-1}}}\over{v_T}}\Big)
          \Bigg[{{D_L}\over{100\, {\rm pc}}}\, (1-x)\Bigg]^{{1}\over{2}}
\end{eqnarray}
The measurable duration of an event is $b_i\, \tau_E$. 

\noindent{\bf Background field:} We 
parametrize the source density in terms of an angle
$\theta_1,$ the average angular separation
between sources in the background field. 

\subsection{Event Rates}
 
The rates of and characteristics of detectable photometric effects of
mesolensing depend on the method of detection and
on the physical parameters of the lens and of the 
background source field. 
The rate of events caused by 
an individual lens of mass $M$, located a distance $D_L$ from us,
lensing a source located at a distance $D_S$ ($x=D_L/D_S$) is 
given by the expression below (Di~Stefano 2006a). In this expression
$\theta_E$ represents the Einstein angle, and $\omega$ is the relative
angular speed between the lens and source. For nearby lenses, $\omega$
is almost exactly equal to the angular speed of the lens with respect to
Earth. 

\begin{eqnarray} 
{\cal R}_{1}^{detect} & = & {{2\, \theta_E\, \omega}
     \over{f_T\, \theta_{mon}^2}}\nonumber\\   
             &   & \nonumber\\  
             &   & \nonumber\\  
             & = & {{0.021}\over{{\rm yr}}}\, \Big({{0.1}\over{f_T}}\Big)\, 
                   \Big({{1''}\over{\theta_{mon}}}\Big)^2\, 
\Big|{{{\bf {\hat v}_L}}\over{100 {\rm km s^{-1}}}} + 
{{{\bf \hat g}}\over{5}}\Big|\nonumber\\
             &   & \nonumber\\
             &   & \times  
                   \Big({{M}\over{0.35\, M_\odot}}\Big)^{{1}\over{2}}
                   \Big({{100 {\rm pc}}\over{D_L}}\Big)^{{3}\over{2}}
                   (1-x)^{{1}\over{2}} \nonumber\\   
\end{eqnarray} 
 
The transverse velocity of the lens is represented by ${\bf {\hat v}_L},$
and the effects of the Earth's motion are incorporated by the vector
${\bf \hat g}$, whose direction depends on position relative to the 
ecliptic and whose magnitude is of order unity. 
The validity of this expression requires that the background field is dense
enough that the monitored region ($\theta_{mon} \times \theta_{mon}$) 
is likely to contain stars.  
Note that the value of $R_1^{detect}$ increases as $\theta_{mon}$ and $f_T$
decrease.  

The total rate of detectable events for a population
of lenses, each with mass $M,$ can be written as follows,
where we assume that $x << 1.\, $ $\Omega_{gal}$ is the surface area of the
galaxy expressed in square degrees. We assume that the density of 
potential lenses is constant. 
\begin{eqnarray}
{\cal R}_{tot}^{detect} & = &\nonumber\\
 &  & {{13.5}\over{\rm yr}}\, {{\Omega_{gal}}\over{\Box^o}}\,
\Big({{0.1}\over{f_T}}\Big)\,
                   \Big({{1''}\over{\theta_{mon}}}\Big)^2\nonumber\\
             &   & \Big({{N_L}\over{0.1 {\rm pc}^{-3}}}\Big)
\Big|{{{\bf {\hat v}_L}}\over{100 {\rm km s^{-1}}}} + 
{{{\bf \hat g}}\over{5}}\Big|\nonumber\\
             &   & \nonumber\\
             &   & \times 
                   \Big({{M}\over{0.35\, M_\odot}}\Big)^{{1}\over{2}}
\Big({{D_L^{max}}\over{{\rm kpc}}}\Big)^{{3}\over{2}}
\end{eqnarray}

\subsection{Light Curve Characteristics} 

It might be supposed that, when the Einstein ring of the
lens is large enough to encompass the positions of several source stars,
multiple-source effects could be important. Indeed, as discussed
below, in rare cases the presence of multiple sources produces
distinctive signals. Nevertheless, most light curves caused by nearby
lenses have the same functional form as light curves caused by more
distant lenses. This is because events in which there is a significant
deviation above baseline tend to be produced by: (a) bright stars;
it is not likely that there are two or more bright stars in any specific
monitored region; (b) close approaches to an individual source star;
it is even less likely that the positions of two bright source stars lie
within a small fraction of an Einstein angle. 

Against very dense source backgrounds, or if the observations are
sensitive enough,   
the event rate can be so high that sequences of events are be
caused by a single lens, particularly if the lens in near enough to
have a proper motion of several $\theta_E$ per year. 
The likelihood 
of sequences of events increases with the value of  
$\theta_{b,i}/\theta_1$.  
Indeed, as the value of $\theta_{b,i}/\theta_1$ becomes as large as
roughly $0.1,$ 
the probability that two or more stars will be detectable lensed
simultaneously increases. 
The simultaneous lensing of two stars produces light curve
shapes that differ from the standard Paczy\`nski form, but which
can nevertheless be well fit by a lensing model in which the 
event is produced by light from a small number of source stars.

\subsubsection{Wide-Field Monitoring Programs} 

The event rate is proportional to the total area monitored.
This means that wide-field monitoring programs can play an important
role in detecting lensing.  
Within the next $5$ to $10$ years two wide-field monitoring projects will
begin. Pan-STARRS, 
{\it Panoramic Survey Telescope and Rapid Response System},
will monitor the sky as seen from Hawaii, (Chambers te al.\, 2004)  and
LSST (the
{\it Large Synoptic Survey Telescope}), 
will operate from Chile (Stubbs et al.\, 2004) .
Together, these surveys will cover the sky. In addition, they will
be able to achieve smaller values of $f_T$ than monitoring programs 
until now. As we will discuss in \S 3,
the combination of wide-field coverage, good photometry,
and high angular resolution will produce high event rates.

\section{Applications: Nearby Stellar Masses}

Nearby stellar populations provide a guaranteed reservoir  
of lensing events. While every nearby star lenses its background,
it is easiest to detect
lensing by objects that are dark or dim in those wavebands
in which the 
background is bright.
Optical monitoring
programs are therefore sensitive to lensing by stellar remnants and
by low-mass dwarfs. These are the lenses we consider in this section.

To compute the expected rate of 
lensing events caused by
nearby stars against any specific background requires
information about the spatial, mass,
luminosity, and velocity distributions of the lenses.
The same types of information are needed for stars in the
source galaxy. The methods employed to detect and select 
events then determine the rate of detectable events.
Given the complexity of the problem, we do not seek to compute rates
with quantifiable uncertainties in this first analysis. 
We can, however, 
use known average properties of the source  and lens
populations, to make reasonable rate estimates by 
substituting appropriately into the formulae derived in paper 1
(Eqs.\, (3) and (4) here).
This allows us to 
answer some general questions.

(1) What are the relative rates of events caused by different types
of lenses? We know, e.g., that some M-dwarf events have been observed.
How many such events must be observed
before we are likely to detect lensing by a T dwarf?  white dwarf?
neutron star? black hole? 

(2) Are the rates of events caused by nearby lenses
 high enough that we 
expect a significant number of
them to be in the data sets of   
(a) completed, (b) ongoing, and (c) future monitoring programs?  

\subsection{The Lenses}

\subsubsection{Stellar Remnants} 
The local spatial density of WDs has been measured,
We use $N_L = 6 \times 10^{-3}$ pc$^{-3}$
(Holmberg, Oswalt, \& Sion 2002; Kawka, Vennes, \& Thorstensen 2004;
Liebert, Bergeron,
\& Holberg 2005). In the calculations
below, we have modeled the WD population with  constant spatial
density equal to this approximate locally-measured value.
This is a reasonable approximation 
because the majority of
existing WDs are descendants of old stellar populations and are
therefore not expected to be confined to a thin disk.
Nevertheless, we expect some
decline in the number of hot young WDs with height above
mid-plane, with thick-disk and halo WD populations becoming
dominant with increasing height. 
If the Galaxy is modeled as a cylinder with radius equal to
$12$ kpc and total height equal to $2$ kpc, a constant density
of $N_L=6 \times 10^{-3}$ pc$^{-3}$  corresponds to a Galactic WD population
of $5 \times 10^{9}$ WDs, with an estimated total mass of WDs of
$3 \times 10^{9} M_\odot.$   We have checked that this number 
is reasonable by comparing it with the results of simulations 
we have conducted.  
Starting with a Miller-Scalo IMF, we 
 find that the total mass in stars is roughly $5-10$ times larger
than the mass in WDs, 
with the ratio depending on the age of the population. Since an estimate of
$3 \times 10^{10} M_\odot$ is consistent with some observational
estimates of the stellar mass of the Milky Way,  
but
low for others, the density employed here does 
not overestimate the WD density. (See, e.g., Sakamoto, Chiba, 
\& Beers 2003; Beers et al.\, 2005.)  
The masses of isolated WDs typically range upward from $0.5\, M_\odot$ 
to close to the Chandrasekhar mass of $1.4\, M_\odot;$
We will use a single value, $0.7\, M_\odot.$  

For NSs we take  $N_L=6 \times 10^{-4}$ pc$^{-3}$, producing 
$5 \times 10^{8}$ Galactic NSs within 1 kpc of midplane.
This is consistent with or lower than most independent estimates
of the size of the NS population (see, e.g., references in Kaspi, Roberts, \&
Harding 2004). We take the mass of NSs to be $1.4\, M_\odot$.
For BHs we take  $N_L=6 \times 10^{-5}$ pc$^{-3}$, producing 
$5 \times 10^{7}$ Galactic NSs within 1 kpc of midplane. (This is
smaller than the numbers often quoted; see, e.g. McClintock \& Remillard 2003).
We take the mass of BHs to be $14\, M_\odot.$  

\subsubsection{Low-Mass Dwarfs} 

The low-mass dwarfs we consider are M dwarfs, L dwarfs, and T dwarfs.
We have drawn the spatial densities for each of these classes, 
shown in Table 1, from the review
by Kirkpatrick  
(2005), with input from Cruz et al.\, (2003),      
Reid, Gizis, \& Hawley (2002),  Knapp (2002), 
Kirkpatrick (2001b), Burgasser (2001), and Gizis et al.\, 2000.   
Note that L dwarfs appear to be relatively rare. This is
because of  both  
the shape of the initial mass function, and 
the fact that the 
transition from stellar to non-stellar objects occurs within the L-range,
and old brown dwarfs will have had time to cool and will therefore 
drop from the class of L dwarfs and will appear instead 
as T dwarfs.     
Naturally each of these classes encompasses a range of masses. 
To simplify the calculations, however, we use a single mass for each class.
For M-dwarf lenses we use $0.2\, M_\odot,$ 
for L dwarfs we use 
$M=0.1\, M_\odot,$ while for T dwarfs 
we choose a sub-stellar mass of $0.05\, M_\odot$.

\begin{table*}
\centering{
\caption{Event Rates and Durations Relative to M dwarfs}
\begin{tabular}{|c||c|c|c|c|}
\hline
Lens type & $\rho$ (pc$^{-3}$) &  Relative value of ${\cal R}_1^{detect} $
                               & Relative value of ${\cal R}_{total}^{detect} $
                               & Relative event duration \\  
\hline
L-dwarfs       &   2e-3   &          0.7    &    0.02    &    0.7  \\ 
T-dwarfs       &   2e-2   &          0.5    &    0.17    &    0.5  \\ 
White dwarfs   &   6e-3   &          1.7    &    0.17    &    1.7  \\ 
Neutron stars  &   6e-4   &         13.      &    0.13    &    0.5  \\ 
Black holes    &   6E-5   &          8.4    &    0.01    &    8.4  \\ 
\hline
\end{tabular}
}
\par
\medskip
\begin{minipage}{0.94\linewidth}
\footnotesize
\end{minipage}
\end{table*}

\begin{table*}
\centering{
\caption{Rates for Different Observing Strategies}
\begin{tabular}{|c||c|c|c|c|}
\hline
  & {\bf Past}  &{\bf Present}& {\bf Future}& {\bf Future}   \\
  & per decade  & per decade  & per decade  & per decade      \\
Lens type&per $\Box^o$&per $\Box^o$&per $\Box^o$&{\bf over 150} $\Box^o$\\
\hline
M dwarfs&(0.3,0.8,2.2)   &(5.7,16,46)    &(110,320,920)&(1.7e4,4.894,1.4e5)\\ 
L dwarfs&(6.4e-3,1.8e-2,5.1e-2)&(0.13,0.38,1.1)&(2.7,7.6,22) &(400,1100,3200)\\ 
T dwarfs&(4.5e-2,0.13,0.36)&(1.0,2.7,7.6)  &(19,54,150)  &(2900,8100,2.3e4)\\ 
WDs     &(5.0e-2,0.14,0.4)&(1.1,3.0,8.6)  &(21,61,170)  &(3200,9100,2.6e4)\\ 
NSs     &(3.6e-2,0.1,0.3)  &(0.76,2.1,6.1) &(15,43,122)&(2300,6400,1.8e4)\\ 
BHs     &(2.3e-3,6.4e-3,1.8e-2)& (4.8e-2,0.14,0.38)&(1.0,2.7,7.7)&(140,400,1200)\\ 

\hline
\end{tabular}
}
\par
\medskip
\begin{minipage}{0.94\linewidth}
\footnotesize
The observing set up for (1) ``past": $f=0.34;$, $\theta_{mon}=2.5'';$ 
(2) ``present": $f=0.1;$, $\theta_{mon}=1'';$
(3) ``future": $f=0.02;$, $\theta_{mon}=0.5''.$
Each prediction consists of a triad of numbers, corresponding to
three values of $D_L^{max}$: ($250$ pc, $500$ pc, $1$ kpc).  
\end{minipage}
\end{table*}

\subsubsection{Flux from the Lenses} 

Among the lenses, the intrinsic optical
luminosity of NSs is low enough that they will not
contribute significantly to light from the monitored region;
the B magnitudes of the known isolated neutron stars tend to be in
the range from 26.6 to 27.2 (Haberl 2005). 
Since isolated black holes have yet to be discovered, we don't know the
range of optical luminosities, but the absence of radiation from
a surface
and the possible advection of energy
through the event horizon suggests that they may be even fainter.     
This is not necessarily true for nearby WDs or cool dwarfs, 
especially if they are young. 
For example, 
cool dwarfs 
have been observed with absolute I magnitudes ranging from 12 to 22
(Dahn et al.\, 2002).
Furthermore, variability has been observed (Bailer-Jones \& Mundt 2001). 
This means that in some cases, light from a nearby lens
may be the dominant source of variable light in the 
wavebands at which it is most luminous. 
When this is the case, present-day algorithms
that select microlensing candidates can fail. 
Therefore, e.g., events caused by some nearby 
cool dwarf stars may not have been identified as lensing candidates.  
For the same reason,  events caused by hot WDs may 
have been identified as lensing candidates only if they were
monitored at relatively 
long wavelengths.

\subsection{Relative rates of detectable events for lenses of different types}

\subsubsection{Individual Lenses} 

M-dwarf-lens events have been detected. 
It is therefore useful
to compute the rates expected for other types of lenses, relative to the rate
for M dwarfs.  
We first compare the rates for individual objects, using Eq.\,~3.
For two objects located the same distance from us, 
and monitored with the same observational set-up, the relative rate depends
only on the ratio of the square root of their masses, and on the ratio
of their transverse speeds. The results are shown in the third column of 
Table 1.

To compute these ratios, we have assumed that the transverse speeds of
all M dwarfs, L dwarfs, T dwarfs, WDs, and BHs are the same.
Observations of pulsars, however, clearly show that the typical speeds of neutron stars
are higher, presumably due to kicks associated with the supernova explosions
that spawned them.    
Although the initial velocity distribution of neutron
stars is highly uncertain,
one widely accepted model is that of Arzoumanian et al. (2002). The
velocity distribution
consists of two Maxwellians. One of these contains $40\%$ of the neutron stars
and is peaked at $127$ km s$^{-1}$; the second, including the rest of the stars,
is peaked at $707$ km s$^{-1}$.)
 $15\%$ of all neutron stars have speeds in excess of $1000$ km s$^{-1}$.
To take into account the relatively higher speed of NSs, we use for them
 a transverse
speed that is five times larger than the transverse speeds used 
for the other stellar lenses.
For this reason, Table 1 indicates that
an individual NS lens can produce an event rate
that is larger than that produced by a BH lens, even though the BH is
more massive.  
The duration of the NS-generated events would tend to
be shorter, however. On average, if we are monitoring a single
lens, the time between events would be shortest for NS-generated events
and longest for T-dwarf-generated events. 
 
\subsubsection{Relative Rates in a Monitored Region} 

If we are monitoring a large region, without 
prior knowledge of individual lenses in the foreground,  
we must use the expression for the total rate in Eq.\, (4), which
includes the relative densities. The results for different stellar
lenses, relative to M dwarfs, are shown in the fourth column
of Table 1. 

These results indicate that the ratio between
the number of M-dwarf events and 
the total number of events caused by all other stellar lenses is
roughly 2 to 1. If, therefore, there are even
a modest number of M-dwarf lensing events in a data set,
the data set is very likely to also include events caused
by compact objects and by brown dwarfs.   
Note that we have assumed that the spatial distributions of these
different classes of lenses are similar. If each population
can be modeled by a spatial distribution that falls off
exponentially with height from the Galactic midplane, and
also
with radial distance from the Galaxy center, then 
differences among the populations are quantifiable
through differences among the exponents applicable for each
population. When these population differences are 
integrated  over a distance of a kpc, relevant for nearby lenses,
they tend to change the relative results by a factor of at most a few.

\subsection{Event Rates} 

Table 2 is designed to predict the rates of events caused by nearby
lenses in optical monitoring programs of the past, present, and future.  
To generate the rates shown in Table 2
we used Eq.\, 4.
We assume three different types of observational set-up.
The set-up labeled ``past" simulates the first generation
of microlensing monitoring surveys. We take $f_T = 0.34$,
and $\theta_{mon}=2.5''$. ``Present" labels the observational
set-up that is similar to the one presently used to monitor M31.
In this case, $f_T = 0.1$,
and $\theta_{mon}=1''$. The moniker ``Future" refers to the type 
of monitoring that will be conducted with Pan-STARRS and LSST.
Specifically, we take $f_T = 0.02$,
and $\theta_{mon}=0.5''$. In the second, third, and fourth
columns of Table 2, we compute the rate per decade per square
degree. In the fourth column, we compute the rate per decade
per $150$ square degrees. This angular area is a conservative
estimate of the area of the sky over which the  combination
of Pan-STARRS and LSST will be able to discover lensing events.

Each entry of Table 2 consists of three numbers. The left-most
number, which is always the smallest, is the rate if $D_L^{max}$ is
$250$ kpc. For the middle 
(rightmost) number, we have used $D_L^{max}=500$ pc 
($D_L^{max}=1000$ pc).    

Table 2 illustrates that improvements in the observational set up,
specifically, lowering the values of $f_T$ and $\theta_{mon}$,
can increase the event rate in a dramatic way. Even if the detection
efficiency for microlensing is low, future monitoring programs
will be capable of detecting large numbers of low-mass stars and stellar
remnants. Existing data provide opportunities to check if
these predictions are overly optimistic.  
We show below that the predictions of Table 2 are consistent
with data 
from the Magellanic Clouds.

\subsection{The Large Magellanic Cloud} 

\subsubsection{Existing Data}
 
The Large Magellanic Cloud has been and continues to be
 monitored by several groups. Here we will focus on the small
set of events identified by the MACHO team through 5.7 years
of monitoring. 
The MACHO team has identified a set of $17$ candidate microlensing
events candidates (Alcock et al.\, 2000).
Of most interest to this paper, $2$ events are now known to have been 
caused by nearby dwarf stars. 

The predictions under the column labeled ``Past" are most relevant
to this data set. Because both M-dwarf lenses were located at
distances close to a kpc, it seems reasonable to take $D_L^{max}$ 
to be roughly a kpc. Using the appropriate entry in Table 2, we
predict $17$ events caused by M dwarfs within a kpc  
($[2.2\, {\rm events\ per\ decade\ per}\ \Box^0] \times  
{5.7{\rm years}} \times  13.5\, \Box^0$).
The efficiency for detecting and identifying these events, {\it if they
are identical with typical microlensing events}, is
roughly $0.3$ (Alcock et al.\, 2000). 
With this value of the efficiency, we expect in the MACHO
data set approximately $5$ identified events caused by
nearby M dwarfs.  
This is consistent with
the two M-dwarf-lens events identified.
The actual efficiency for detecting lensing events caused by
nearby lenses is likely to be even lower, because some nearby
lenses may have been bright enough to produce chromatic
effects which would have eliminated the events from consideration. 
   It is certainly possible that additional M-dwarf events have been 
detected, but are not yet identified as promising lensing candidates.
In addition, the data set could contain one or two events caused by
either a compact stellar remnant or a T dwarf.

\subsubsection{Future Observations}

When LSST monitors the LMC, the parameters used in Table 2
 for ``future" observations
will apply, at least to a large portion of the LMC.
Specifically, LSST (1) will likely achieve values of $f_T$
smaller than $0.02,$ and, (2) in a single observation, 
it will reach limiting
magnitudes dimmer than 24, so that regions $0.5''$ on a side will
contain mulitple stars that can be detectably lensed. 
Considering only
the portion of the LMC that was monitored by the MACHO team, we 
multiply the values for $D_L = 1$ kpc in 
column 4 (``future") of Table 2 by $13.5$, to predict that
in this region alone, a decade of LSST observations will detect 
$1.2\times 10^4$ M dwarf events, with another 
$\sim 6 \times 10^3$ 
events caused by other dwarfs
and by stellar remnants, including
$\sim 1600$ neutron star events and $100$ black hole events. 

It is interesting to note that, even if we make very conservative
assumptions, the numbers of events generated by nearby lenses in
future monitoring programs is guaranteed to be high. 
If, for example,
 the effective value of $D_L$ is actually only $250$ pc,
four times smaller than seems consistent with the
MACHO results,
this would decrease the rate by a factor of only $8$. If,
in addition, the efficiency is
$10\%$, which is lower than expected, nearby lenses
 would still produce a large
number of events. For example, under these conservative assumptions,
this limited portion of the LMC
should, over a decade, exhibit  $20$ neutron-star generated events 
($150$ M-dwarf-generated events),   
each potentially measuring mass and multiplicity.  
Considering the entire sky, the numbers increase by a factor
of at least $10$. 
Although assumptions like the ones we make in
this paragraph are inconsistent with the MACHO LMC and OGLE Bulge results,
and therefore lead to unrealistically 
small lower limits on the numbers of expected events,  
they nevertheless predict a significant number of mesolensing events.

Considering the LMC itself, 
area out to the D25 isophote is 6-7 times larger 
than the area monitored by the MACHO team. Nevertheless, the total
event rate will not be larger by a factor this large because the 
area that will be added by LSST is generally less dense than the area
originally monitored. This means that the effective value of $\theta_{mon}$
must be larger. With, e.g.,
 $\theta_{mon} \sim 3''$, corresponding to roughly
65 stars in each the monitored region for a stellar density of
$0.1$ pc$^{-3}$, the added regions of the LMC would
contribute roughly $1/6$ as many events as the central region.
While the central region would dominate, the number of events
detected against the outer portions of the galaxy
is also large.

\subsubsection{Disk of M31} 

M31 has been the subject of regular monitoring
by independent teams during the 
past several years. Image differencing methods have been
used to identify candidate microlensing
events, using observing parameters $f_T$ and $\theta_{mon}$
roughly comparable to those used in the row labeled ``Present"
in  
Table 2.
The surface area of M31 lying within the D25 isophote is
approximately a square degree. In one year of monitoring,
we therefore predict $(1, 2.4, 7.2)$ events if
$D_L^{max}$ is ($250\, {\rm pc}, 500\, {\rm pc}, 1\, {\rm kpc})$. 
If the effective total
duration of past observations is on the order of a year, 
it is possible that some of the events detected have been
caused by nearby lenses; direct detection of the lens,  
 or a pattern of continued time variability consistent with sequential
lensing of different source stars could provide definitive
evidence.
Given the crowding of stars in M31, and
the photometric sensitivity of current observations, we
expect that $b_i$ for typical M31 events should be small,
perhaps $\sim 0.1.$ This means that event durations will be
shorter than is typical against the much less crowded background of
the Magellanic Clouds.  
It is therefore interesting that a large fraction of the
candidate microlensing events are short, with 
durations as measured by the full-width half maximum ($t_{FWHM}$)
smaller than $5$ days.
Paulin-Henriksson et al.\, (2003) report on
$4$ events, $3$ of which have $t_{FWHM} < 2.2$ days;
the fourth has $t_{FWHM} = 21.8$ days.
de~Jong et al.\, (2004) report on $14$ microlensing candidates,
$5$ of which have $t_{FWHM} < 5$ days, one with $t_{FWHM} = 9.4$ days,
and an additional six candidates with $t_{FWHM} < 30$ days.
(See also Belokurov et al.\, 2005.)

\noindent{\bf Future Observations:} More 
significant in the long term, however, is the potential
of future programs to detect lensing against the background of M31.
Pan-STARRS will monitor M31. In a decade of monitoring, 
data from 
$\sim 500^{+900}_{-300}$ events caused by nearby lenses should be collected. 
Even if the detection efficiency is
only a few percent, lensing by more than a dozen M dwarfs will be
identified, and a neutron star could be discovered.
The higher event rate for lenses of all types is caused by the lower
values of $\theta_{mon}$ and $f_T$ that are possible with 
Pan-STARRS.

\noindent{\bf The Halo of M31:} The  
wide-field element of Pan-STARRS will also be important
because it extends monitoring to the halo of M31. 
All other things being equal, the rate of events is proportional
to the surface area of the background source field. The surface area of the
halo of M31 is $\sim 100$ times larger than the surface area of
the disk. On the other hand, the average density of stars in the halo
is lower than in the disk.
For low stellar densities, the rate of events can be computed
by considering the rate at which bright background stars enter
the Einstein ring of the lens. (See paper 1, Di\thinspace~Stefano 2005.)
But it is also possible to use Equations (3) and (4) of this
paper, as long as the value of $\theta_{mon}$ is chosen
appropriately. 

In order for the  formalism that was used to derive these
expressions to be applied, regions of size $\theta_{mon}$
must, on average, contain one or more source stars bright
enough to produce a detectable baseline. For dense fields,
the smallest such region is smaller than the smallest region
the observing teams can monitor, based on present-day spatial
rsolution and pixel sizes.
That is, we are
limited by present-day technology. For low-density fields,
on the other hand, 
achievable values of $\theta_{mon}$ may be too small for
a field of size $\theta_{mon} \times \theta_{mon}$
to contain bright stars. This 
doesn't mean that observers should degrade their resolution.
Instead, if we want to use Eqns. (3) and (4) to compute
event rates, we must use an effective value of $\theta_{mon}^{eff},$
large enough to contain some source stars, and possibly larger than the actual values of $\theta_{mon}$ to be used in
observations.   
 The dependence of the rate on the source 
surface density then enters through the factor of 
$1/\theta_{mon}^2$, which should be replaced by 
$(\theta_{mon}^{eff})^2$.
A lower limit on $\theta_{mon}^{eff}$ can be estimated by first  
computing $\theta_1,$ where $\sigma_S \theta_1^2 = 1.$ 
In the disk of M31, if $N_S$ is approximately $0.1$ pc$^{-3}$,
then $\theta_1\sim 0.027''.$ If the monitored field must encompass
roughly $100$ ($10$) stars in order for one of them to be bright enough to
be detectably lensed, then $\theta_{mon}^{eff} = 10\, \theta_1 \sim 0.27''$
($\theta_{mon}^{eff} = 3.2\, \theta_1 \sim 0.08''$).
This verifies that, in the disk of M31, the lower limit
on the  value of $\theta_{mon}$ is
at present determined by technical considerations, not by the density of stars.
As we move outward from the center of the galaxy,
we can continue to
use values of $\theta_{mon} = 0.5''$ ($\theta_{mon}$ of $1''$) in regions with
average stellar densities as low as $25\%$ ($6\%$) of the average disk density.
This general argument simply points out that there can be a  
region surrounding the disk of the galaxy 
in which the rate of lensing events can be comparable to the rate
in the disk. 

Farther from the center of the disk, the value of $\theta_{mon}$
to be used in the calculations, $\theta_{mon}^{eff}$, 
is determined by the local stellar surface density. 
If the average surface density of the major portion of the halo is 
$0.0001$ stars pc$^{-3}$, $\theta_{mon}^{eff}$ would be $\sim 8.4''$, for
regions containing $100$ stars.  
Those source stars that can be detectably lensed in the halo
will consist primarily of subgiants and giants. If  
this area of the halo covers an area equal to $100$ times the area
of that part of the disk with $\theta_{mon} = 0.5'',$ then the 
rate of halo events will be $0.35$ the rate of disk events. 

Lensing of halo sources will produce hundreds of
events per decade. The spatial distribution of the events, 
and the range of time durations as a function of position will 
map the density and
luminosity distribution of halo stars. 

\subsubsection{More Distant Galaxies}

As we look toward the disks of more distant galaxies,  
the intrinsic rate per lens
is the same as for the disk of M31. This means that the total
rate of detectable events scales as the projected area of the galaxy.
At $10$ Mpc, the area per galaxy is smaller by a factor of $100$. 
On the other hand, if homogeneity is roughly correct, then an argument analogous to the one made in Olber's
paradox, shows that the number of galaxies in equal-width shells 
at larger distances from us increases in just such a way as to keep
the total area of sky (per shell) covered by galaxies the same.
Our ability to detect events against the background of more distant
galaxies is limited only by the facts that (1) the greater distance
makes source stars in more distant
galaxies appear fainter,  while (2) there is a cut-off at the high end
of the luminosity function. As we consider 
background galaxies at greater distances, the 
larger apparent magnitudes of even the brightest stars  
require smaller and smaller values of $b_i$.
Consequently, the 
time durations of events decrease until they become impossible to observe.

Nevertheless, there is a large volume around us
within which galaxies provide useful backgrounds for lensing 
by nearby stars.  
For example, the disks of galaxies other than
M31 which happen to be located within $5$ Mpc, but farther from us
than $0.7$ Mpc cover an approximately $3$ square degrees of the sky
(Tully 1988).  
In an all-sky monitoring program, the rate for the main body (excluding
the halos) of external galaxies, could contribute
several times more events than 
M31.

Just as the halo of M31 provides a background against which
events can be detected, so do the halos of other
external galaxies.
As the distance from us increases, the ratio of the rate of lensing
associated with the halo to that associated with the main
disk of its galaxy will decrease, at least for
late-type
galaxies. This is because the brightest halo
stars are likely to be low-mass giants, and they tend to be less 
luminous on average than the most luminous stars in
young stellar populations. 

\subsubsection{The ICM of Galaxy Clusters} 

If a cluster at $20$ Mpc has a
radius of $1$ Mpc, it subtends an area on the sky of roughly 
$25$ square degrees. Most of that area is filled with the 
intracluster medium (ICM).    
Although not as dense as galaxies, the ICM can be a rich
environment.
The total light emitted by Virgo's ICM, e.g., is estimated to be
$10-20\%$ of the light emitted by the cluster's galaxies (Feldmeier et al.\,
2004a, 2004b;
see also Arnaboldi 2003, 2004). Comparing the ICM to a galaxy, 
there may be $10-20\%$
as many stars in a volume (surface area) $1000$ ($100$) times greater. 

The role of the ICM with respect to the galaxies in a cluster
is analogous to the role played by the halo of M31 relative to the
galaxy's disk. 
It is interesting that the distribution of stars in parts of
Virgo's ICM has been observed to exhibit significant structure.
{\it Over the long term, mesolensing observations could play a role in
helping to map such structure by, e.g., studying
how the event durations change with position.} 

\subsubsection{All-Sky Monitoring}

The calculations above indicate that lensing events caused
by nearby stars are present in existing data sets, and will be
found in large numbers 
in the data sets collected by sensitive all-sky monitoring programs,
such as Pan-STARRS and LSST. The durations will tend to larger than
a few days,  
so the events will be discovered and the longer ones will
be reasonably well sampled
by the monitoring programs.  

Lensing will therefore 	
become a tool to discover very dim nearby stellar-mass and planet-mass
objects. Under ideal circumstances
lensing can also measure the masses of these objects,
their proper motions, and provide information about multiplicity. 
This will lead to an unprecedented amount of information
about stellar remnants and low-mass dwarfs, 
even should detailed measurements be possible for only a subset
of the events, e.g., those which are studied
with ``follow-up" observing programs that 
use more sensitive detectors and a higher
sampling frequency. 

As events are discovered over a large portion of the sky, we will
learn about the background as well as about the lenses. This is
because lensing can be detected across a wide range of
background stellar densities. The spatial distribution of events
and the distributions of event properties, especially the durations, 
can be linked to the density of the background field and
can therefore be used to map the background.

\section{Applications: Possible Lenses}

In this section the focus shifts to potential lenses 
whose existence has not been established. 
In these cases, mesolensing may provide a way to test for the presence
of these objects and to either discover them or to derive
upper limits on their populations.

\subsection{Intermediate-Mass Black Holes}
An increasingly active field of research is focused
on so-called intermediate-mass BHs (IMBHs), with possible masses between roughly
$50$ and $10^5$ solar masses (see, e.g., Miller \& Colbert 2004). 
Although there is no direct evidence for IMBHs, increasing numbers of
ultraluminous X-ray sources (ULXs) are being discovered in external
galaxies (see, e.g., Mushotzky 2004). 
With X-ray luminosities typically between $10^{39}$ erg s$^{-1}$
and $10^{41}$ erg s$^{-1}$, they are super-Eddington for 
$10\, M_\odot$ objects, suggesting that more massive compact objects,
almost certainly BHs, may exist. If this interpretation of
the evidence is correct, then the IMBHs in ULXs must represent
a tiny fraction of all IMBHs. Most IMBHs may not be in binaries,
and those that are may not be X-ray active at present. Those
that are X-ray active are not likely to be in a ULX state.
Detailed population estimates have not yet been made.
The presence of IMBHs is also consistent
with theoretical considerations suggested by the presence of
supermassive BHs in galaxy centers 
(see, e.g., Volonteri, Madau,
 \& Haardt 2004; Islam, Taylor, \& Silk 2004 and 
references therein) and by the evolution of massive
young star clusters (see, e.g., Portegies 
Zwart et al.\, 2004; Freitag et al.\, 2005 and
references therein). If IMBHs exist, can they be observed against the 
background of M31 or  other stellar fields?

If there are $10^7$ IMBHs in the Galaxy, then, if the Galactic volume is
approximately $300\, {\rm kpc}^3,$ there is roughly $1$ IMBH 
per kpc$^3$ per square degree.
This corresponds to a spatial density of 
roughly $3\, \times 10^{-6}$ pc$^{-3}$. 
There may therefore be one IMBH lying in front of the main disk of
M31, $\sim 20$ lying over the densest region of the LMC, $\sim 100$
in total in front of the LMC, and at least $\sim 200$  in front of the
regions in which Pan-STARRS and LSST will find the most lensing events.

The question is then, whether we are likely to have detected lensing by IMBHs.
Consider a lens with mass $M=1400\, M_\odot.$ 
If there are no other modifications to Eq.\, (3), then
${\cal R}_1^{detect}$ is $1.33$ yr$^{-1}$. The true rate will be lower however,
because $D_L$ will, on average, be greater than $100$ pc.
Taking $D_L$ to be $500$ pc, we find ${\cal R}_1^{detect}=0.12$ yr$^{-1}$.
Even this is likely to be too optimistic, since the appropriate value of
$f_T$ for existing data is more likely be $\sim 0.34,$ rather than $0.1;$
this reduces the value of ${\cal R}_1^{detect}$ to 
$3.5\, \times 10^{-2}$~yr$^{-1}$. The event time duration (with
$D_L=500$ pc is $b \times 6.8$~yr.  This resulting duty cycle
yields a $24\%$ chance that any individual IMBH will serve as a lens. 
With, e.g., roughly a dozen IMBHs lying in front of the LMC, and a similar
number in front of the Bulge along Baade's window, there is a chance that
lensing by an IMBH has been observed.

The probability of detecting lensing caused by an IMBH will be
significantly increased with the advent of Pan-STARRS and
LSST. With $f_T$ decreased to $\sim 0.02,$ the value of
${\cal R}_1^{detect}$ increases from $3.5\, \times 10^{-2}$~yr$^{-1}$
to $0.6$~yr$^{-1}$. The time duration of events, which can then be followed
to magnifications as small as $1.02,$ increases to $b\times 19.3$~yr.
It therefore seems likely that an individual light curve will 
contain simultaneous signals of the ongoing lensing of several
source stars. The signal will be complicated, but should be well-fit
by lensing models.  Failure to detect an appropriate signal
with Pan-STARRS and LSST will allow us to place limits on the density of
IMBHs with masses near $1000\, M_\odot$ of $10^{-8}-10^{-7}$~pc$^{-3}$,
and comparable limits even on BHs with masses of $\sim 100\, M_\odot.$

Once a region containing a possible IMBH lens is identified, it is 
possible that high-angular-resolution measurements will allow
spatial effects associated with the lensing to
be studied. In some cases (for the closest IMBHs
or for those with masses $> 10^3 M_\odot$), the Einstein ring may be $\sim 0.1''$,
making astrometric effects more accessible to measurement.

\subsection{Free-Streaming Dark Matter in the Solar Neighborhood}

The microlensing monitoring programs were able to place significant
constraints on the fraction of the Halo comprised of
compact objects in a variety of mass ranges. Below we show
that monitoring
programs should be able to
detect or place limits on the presence of nearby compact objects.
We don't know the underlying mass distribution of such
objects, should they exist. We can, however, make assumptions about
the range of possible masses, and determine which
mass values would lead to detectable lensing signatures.

If, e.g., local dark matter has a density of
$9 \times 10^{-25}$ gm cm$^{-3}$ (Gates, Gyuk, \& Turner 1995), there could be roughly
$100$ Jupiter mass ($M=10^{30}$ gm) objects within a pc of Earth.
Their Einstein angles can be small. 
\begin{equation}
\theta_E = 0.002'' \Bigg[{{M}\over{10^{30}{\rm gm}}}\, {{{\rm pc}}\over{D_L}}\, 
(1-x)\Bigg]^{{1}\over{2}}
\end{equation}
Nevertheless, finite source size effects should not
interfere with detection when $D_S$ is large. Indeed, $\theta_E$ would have to be
$6 \times 10^{-7}{''}$ in order for the Einstein ring at the distance to M31
to be as small as $100\, R_\odot.$  A more serious concern is finite lens size
effects. Although dark matter may be transparent, the calculations will be altered
if the mass distribution of the lens must be explicitly considered.
If, however, the dimensions of the dark matter are similar to
those of planets, finite lens size effects will be rare. 
[See Eq.\, (10).] 
 
If these dark masses exist and if they are freely streaming through the Galaxy with
typical velocities of $200$ km s$^{-1}$, the angular velocities
of those nearby can be large, as can the rate of events generated
by a single such lens.
\begin{equation}
\omega={{42''}\over{\rm yr}}\, \Big({{v}\over{200\, {\rm km/s}}}\Big)\,
\Big({{\rm pc}\over{D_L}}\Big)
\end{equation}
\begin{eqnarray}
{\cal R}_{1}^{detect}& = & {{1.7}\over{\rm yr}}\, 
\Big({{0.1}\over{f_T}}\Big)\,
                   \Big({{1''}\over{\theta_{mon}}}\Big)^2\,
\Big|{{{\bf {\hat v}_L}}\over{200 {\rm km s^{-1}}}} +
{{{\bf \hat g}}\over{10}}\Big|\nonumber\\
             &   & \nonumber\\
             &   & \times
                   \Big({{M}\over{10^{30}{\rm gm}}}\Big)^{{1}\over{2}}
                   \Big({{1 {\rm pc}}\over{D_L}}\Big)^{{3}\over{2}}
                   (1-x)^{{1}\over{2}} \nonumber\\
\end{eqnarray}
The time duration per event is small.  
\begin{eqnarray} 
\tau_E &=&{{2\, \theta_E}\over{\omega}} \nonumber\\ 
       &=&48\, {\rm min}\, \Big({{M}\over{10^{30} {\rm gm}}}\Big)^{{1}\over{2}} 
          \Big({{200\, {\rm km\, s^{-1}}}\over{v_T}}\Big)
          \Bigg[{{D_L}\over{1\, {\rm pc}}}\, (1-x)\Bigg]^{{1}\over{2}}      
\end{eqnarray}

Note that, if deviations from baseline of $2\%$ can be
detected, the duration of an event can be $3$ times longer--i.e.,
the time taken to cross $3$ Einstein diameters.
Given the fact that there is a distribution of velocities, and also
a distribution of orientations, producing an even broader
distribution of transverse velocities, we expect both
shorter and longer events,

Because the Einstein ring
is small, we are in the low-density regime, and
events should primarily be single-source events.
Nevertheless, as Eq.\, (7) indicates, 
sequences of
events are expected because the angular speed is large.  

The total rate per square degree is also high.
Within $100$ pc, there would be $\sim 100$ masses of $10^{30}$ gm 
 per square degree.
\begin{eqnarray}
{\cal R}_{tot}^{detect} & = &\nonumber\\
 &  & {{1130}\over{\rm yr}}\, {{\Omega_{gal}}\over{\Box^o}}\,
\Big({{0.1}\over{f_T}}\Big)\,
                   \Big({{1''}\over{\theta_{mon}}}\Big)^2\nonumber\\
             &   & \Big({{N_L}\over{110 {\rm pc}^{-3}}}\Big)
\Big|{{{\bf {\hat v}_L}}\over{200 {\rm km s^{-1}}}} +
{{{\bf \hat g}}\over{10}}\Big|\nonumber\\
             &   & \nonumber\\
             &   & \times
                   \Big({{M}\over{10^{30} {\rm gm}}}\Big)^{{1}\over{2}}
\Big({{D_L^{max}}\over{{\rm kpc}}}\Big)^{{3}\over{2}}
\end{eqnarray}

The duty cycle for events associated with these
nearby lenses could be high enough to allow monitoring specifically
designed to discover short-duration events to either discover or place
limits on planet-mass, free-streaming dark matter. 

If short-duration events are observed, and if  
the monitoring occurs regularly over the period of several years,
several tracks of events should be observed. 
The number of tracks actually associated with lenses
effect can be estimated through statistical simulations.
Individual cases in which the tracks are reliably established 
as being caused by lensing will be important, because the  
distance to the
lens and the lens may may be measured in such cases. 

If no such events are discovered, 
limits on disk  dark matter in the form of Jupiter-mass objects
can be derived. Such limits 
on Halo dark matter are already strong,
based on the EROS and MACHO data Alcock et al.\, 1998.
Mesolensing observations, however, can place limits on disk dark matter.
In addition, the limits can be placed on lower mass values.  
For lower-mass objects, the Einstein angle is smaller. This does not cause
problems with finite source size, but it does decrease the time of events,
making it more difficult to discover them. If, however, all of the dark matter
is in the form of such low-mass objects, there will be more of them.
This means that a larger number will be closer, so that the distribution of
values of $\theta_E$ will still include some larger values. It also means
that there will be a larger number of lenses with velocities directed along
more radial paths, producing fewer events per year, but with each event
having a longer duration than if the motion were perpendicular to
our line of sight.

\subsection{Low-Mass Planets in the Outer Solar System}

We don't know whether the outer regions of our solar system,
harbor planet-mass
objects. 
At distances of $\sim 1000$ AU a planet would have to be far more
massive than
Jupiter, in order for its dynamical influence on the
known outer planets to be discernible (Hogg et al.\, 1991).  
Mesolensing provides an independent way to derive limits.

It is possible, if a planet-mass object is close enough to us,
for its angular size 
to be larger than its Einstein ring.
The requirement that it fit inside
its Einstein ring is:
\begin{equation} 
{{D_L}\over{1000\, {\rm AU}}} > 0.11\, \Big({{M_J}\over{M_p}}\Big)
                                \Big({{R_p}\over{10^9 {\rm cm}}}\Big)^2,   
\end{equation} 
where $M_p$ and $R_p$ are the mass and radius of the planet, respectively.  
harbor planets comparable in mass 

For a Jupiter-mass planet located $2000$ AU from us, the
size of the Einstein ring is roughly $0.02''.$
For less massive planets, however, the Einstein angle can
be in the milliarcsecond range (e.g., the Earth at $2000$ AU)  or even smaller.
The event rate is high nevertheless,
because the angular speed on the sky is high.
A mass $2000$ AU from us traverses an angle of approximately $200''$
every $6$ months. If the Einstein angle is $0.02'',$ the lens will
move through $\sim 5000$ Einstein diameters in $6$ months, and the 
Einstein crossing time will be approximately an hour.
The area we will perceive the lens to cover across the source field
per year is large: $4\, \Box''$ every $6$ months.

The background field could be out of the ecliptic, because  
the distribution of outer solar system masses 
may be close to spherical. If such a planet, or planet exists, 
lensing by it is therefore most likely to be detected by wide-field
monitoring programs like Pan-STARRS and LSST. The smaller values
of $f_T$ possible for these programs will also increase the
rate at which any such planets cause events.  
Although the frequency of detected events and their durations will depend on
the direction, there are important common elements across directions.
First, events will be of short duration, typically hours or less.
Second, individual lenses should each give rise to sequences of 
events. These sequences will trace the motion of the Earth,
exhibiting a ``forward" and ``backward" motion every year. This
back and forth swing will repeat on a yearly basis, 
with slight changes due to the
relatively slow motions of the lens and background stars.      
It is important to note that astrometric effects can be 
significant as well, and may be studied with, e.g., Gaia
(see Gaudi and Bloom 2005).

\section{Conclusion}

\subsection{Guaranteed Signal}

Discussions of ``dark matter" typically focus on matter whose nature is
yet to be understood.
The local neighborhood is,
however, filled with dark objects whose nature we think we understand
(BHs, NSs, WDs, and low-mass dwarfs), but of which we have relatively
few nearby examples. Mesolensing provides a way to conduct a census
of such objects,
providing mass estimates and distances for many, 
 and opening the door to detailed study of some
individual systems.  
Mesolensing observations are guaranteed to identify
BHs, NSs, WDs, and low-mass stars within $\sim 1$ kpc of Earth.  

The calculations of \S 3 indicate that events due to nearby
lenses 
(1) should be part of virtually every microlensing data
set collected so far, and that  
(2) wide-field monitoring programs will regularly discover
signals due to nearby lenses; this provides additional motivation
for Pan-STARRS and LSST. 
(3) A combination of programs will regularly
discover BHs within a few kpc, as well as NSs and other objects
of lower-mass.  

\subsection{Tests for Exotic Matter} 
Mesolensing can also discover or place limits on the existence
of disk dark matter, and can contribute
to surveys of the outermost regions of the solar system. 
Under certain circumstances it can be used to study 
more distant masses in binaries, possibly
including intermediate-mass black holes in
our own Galaxy and the supermassive black holes found at the 
centers of galaxies (Di\,~Stefano 2007b).

\subsection{Pan-STARRS, LSST, and other monitoring programs} 

One of the interesting results we have derived is the high 
rate of events possible with the coming generation of
monitoring programs. Key factors in increasing the rate
 are the fact that
sensitive photometry will allow smaller values of $f_T,$ 
while better spatial resolution will allow 
smaller values of $\theta_{mon},$ at least in fields
crowded enough to contain stars in a 
surface area $\theta_{mon}^2.$
This means that the rate of events due to nearby stars
per square degree will be significantly higher than at present.  
Pan-STARRS and
LSST also naturally benefit from being able to monitor large
areas. We  have used what appear to be realistically
achievable values of the parameters $f_T$ and $\theta_{mon}$.
Even if the values eventually achieved are not optimal,
the predicted rates are still high. The detection
efficiency then determines the fraction of events that
will be successfully identified.  
Detection efficiencies for past monitoring programs have
typically been in the range of tens of percent. It is
likely that new  monitoring programs will be at least 
comparably efficient. Even though the cadences of wide-field
monitoring may not be ideal for lensing, the greater
photometric sensitivity means that events will typically last
longer, providing more opportunities for detection.
The upshot is that the predicted rate of events caused by
nearby stars will be high enough to ensure that many
nearby lenses will be detected, and some will be studied
in detail. 

It is interesting to note that, in general, we expect
lensing events to be caused by nearby lenses, lenses in the
background source field (``self-lensing'') and possibly
by MACHOs. Although our calculations were carried out
for nearby lenses, the same considerations
hold for other lenses as well, at least for a wide
range of background fields. It is 
therefore  clear that future  monitoring programs, especially
Pan-STARRS and LSST  will be important sources of lensing
data.     
   
\bigskip
\bigskip
{\footnotesize     



\noindent
Afonso, C., et al.\ 
2003, A\&A, 404, 145 

\noindent Afonso, C., et al.\ 
2000, \apj, 532, 340 

\noindent  Alard, C., \& Lupton, R. H. 1998, ApJ, 503, 325 

\noindent
Alcock, C., et al.\ 
2001, \apj, 552, 259 

\noindent
Alcock, C., et al.\ 
2001, \apj, 562, 337 

\noindent Alcock, C., et al.\ 
2000, \apj, 542, 281 

\noindent
Alcock, C., et al.\ 
1998, \apjl, 499, L9 
\noindent
Alcock, C., et al.\ 
1995, \apjl, 454, L125 

\noindent Arnaboldi, M.\ 2003, Memorie 
della Societa Astronomica Italiana Supplement, 3, 184 

\noindent
 Arnaboldi, M. et al.\ 2004, \apjl, 614, L33

\noindent
Arzoumanian, Z. , Chernoff, D. F., \&  Cordes, J. M. 2002, ApJ, 568, 
289A

\noindent
Bailer-Jones, 
C.~A.~L., \& Mundt, R.\ 2001, A\&A, 367, 218

\noindent
Beers, T.~C., et al.\ 
2004, IAU Symposium, 220, 195 

\noindent
Belokurov, V., et 
al.\ 2005, \mnras, 357, 17 


\noindent
Burgasser A. J. 2001; 
The discovery and characterization of 
methane-bearing brown dwarfs and the definition of the T 
spectral class. PhD thesis. Calif. Inst. Technol.
of 21E-3/pc3 for the T5-T8 range using 14 T dwarfs identified from 2MASS.

\noindent
Caballero, J.~A., 
B{\'e}jar, V.~J.~S., Rebolo, R., \& Zapatero Osorio, M.~R.\ 2004, A\& A, 
424, 857 

\noindent
Chambers, 
K.~C., \& Pan-STARRS 2004, American Astronomical Society Meeting Abstracts, 
205,  

\noindent
Cruz K.L., Reid I.N., Liebert J., Kirkpatrick J.D., Lowrance P.J.\, 2003, 
Astron. J. 126:2421-48
 
\noindent
Dahn, C.C. et al.\,  2002, AJ, 124, 1170 

\noindent
de Jong, J. T. A., et al. 2004, A\& A, 417, 46

\noindent
Di\thinspace Stefano 2005a, {\it Binary Mesolenses}, in preparation 

\noindent
Di\thinspace Stefano 2005b, {\it Lensing Tests for Supermassive 
Black Holes}, in preparation 

\noindent
Di\thinspace Stefano \& Kong 2005, {\it X-Ray Sources in the Halo of M31,}
submitted.

\noindent
Dominik, M., \& Sahu, 
K.~C.\ 2000, \apj, 534, 213 

\noindent 
Drake, A.~J., Cook, 
K.~H., \& Keller, S.~C.\ 2004, ApJL, 607, L29 

\noindent
Dyson, F.W., Eddington, A.S., \& Davidson, C. 1920, Philos. Trans. R. Soc.\,
London, A220, 291

\noindent
Eddington, A.~S., 
Jeans, J.~H., Lodge, O.~S., Larmor, J.~S., Silberstein, L., Lindemann, 
F.~A., \& Jeffreys, H.\ 1919, MNRAS, 80, 96 

\noindent
Feldmeier, J.~J. et al\ 2004a, \apj, 615, 196 

\noindent Feldmeier, J.~J.\ 2004b, 
 astro-ph/0407625 

\noindent
Freitag, M., Atakan 
G{\" u}rkan, M., \& Rasio, F.~A.\ 2005, 
astro-ph/0503130 

\noindent
Gates, E.~I., Gyuk, G., 
\& Turner, M.~S.\ 1995, \apjl, 449, L123 

\noindent
Gaudi, S. \& Bloom, J. 2005.

\noindent
Gizis J.E., Monet D.G., Reid I.N., Kirkpatrick J.D., 
Liebert J., Williams R.J.\, 2000. Astron. J. 120:1085-99  

\noindent
Gould, A., Bennett, 
D.~P., \& Alves, D.~R.\ 2004, ApJ, 614, 404 

\noindent Gould, A.\ 2004, ApJ, 606, 319 

\noindent Gould, A.\ 1996, \apj, 470, 201 

\noindent
Griest, K.\ 1991, ApJ, 366, 
412 

\noindent
Haberl, F. 2005, MPE Report Vol. 288, 39; astro-ph/0510480

\noindent
Hogg, D.~W., Quinlan, 
G.~D., \& Tremaine, S.\ 1991, \aj, 101, 2274 

\noindent
Holberg, J.~B., Oswalt, 
T.~D., \& Sion, E.~M.\ 2002, \apj, 571, 512 

\noindent
Islam, R.~R., Taylor, 
J.~E., \& Silk, J.\ 2004, \mnras, 354, 629 

\noindent
Jiang, G., et al.\ 2004, 
ApJ, 617, 1307 

\noindent 
Kirkpatrick J. D. 2001b; In Tetons 4: Galactic 
Structure, Stars and the Interstellar Medium, ed. CE Woodward, MD Bicay, JM Shull, ASP Conf. Ser. 231:17-35

\noindent 
Kirkpatrick, J.~D.\ 2005, 
ARA\&A, 43, 195 

\noindent
Kochanek, C.~S.\ 2004, ApJ, 
605, 58 

\noindent
Kaspi, V.M., Roberts, M.S.E., Harding, A.K.  
2004, to appear in "Compact Stellar X-ray Sources", eds. W.H.G. Lewin and M. van der Klis,
Cambridge University Press,
astro-ph/0402136.

\noindent
Kawka, A., Vennes, S., \& 
Thorstensen, J.~R.\ 2004, \aj, 127, 1702 

\noindent
Kerins, E., et al.\ 
2003, \apj, 598, 993 

\noindent
Kleinman, S.~J., et 
al.\ 2004, \apj, 607, 426 

\noindent 
Knapp, G.~R.\ 2002, Bulletin of
the American Astronomical Society, 34, 1215

\noindent
Liebert, J., Bergeron, 
P., \& Holberg, J.~B.\ 2005, \apjs, 156, 47 

\noindent Luhman, K.~L., Fazio, 
G., Megeath, T., Hartmann, L., \& Calvet, N.\ 2005, Memorie della Societa 
Astronomica Italiana, 76, 285 

\noindent
Luhman, K.~L.\ 2004, \apj, 616, 
1033

\noindent
Luyten, W.~J.\ 1999, VizieR 
Online Data Catalog, 3070, 0 

\noindent
McClintock, J.E. \& Remillard, R.A. 2003, 
to appear in "Compact Stellar X-ray Sources," eds. W.H.G. Lewin and M. van der Klis, 
Cambridge University Press, astro-ph/0306213 

\noindent
McCook, G.~P., \& Sion, 
E.~M.\ 1999, ApJS, 121, 1 

\noindent
Mao, S., et al.\ 2002, 
MNRAS, 329, 349 

\noindent
Mei, S., et al.\ 2005, 
ApJS, 156, 113 

\noindent
Miller, M. C., \& Colbert, E. J. M. 2004, Int. J. Mod. Phys. D, 13, 1

\noindent 
Mohanty, S., 
Jayawardhana, R., \& Basri, G.\ 2004, ApJ, 609, 885 

\noindent
Mushotzky, R.\ 2004, 
Progress of Theoretical Physics Supplement, 155, 27 

\noindent
Nale{\. z}yty, 
M., \& Madej, J.\ 2004, A\&A, 420, 507 

\noindent
Nguyen, H.~T., 
Kallivayalil, N., Werner, M.~W., Alcock, C., Patten, B.~M., \& Stern, D.\ 
2004, ApJS, 154, 266 

\noindent
 Okamura, S., et al.\ 
2002, \pasj, 54, 883 

\noindent
Paczy\'nski, B.\ 1996, ARAA, 
34, 419 

\noindent
Paczy\'nski, B.\ 1986, ApJ, 
304, 1

\noindent
Paulin-Henriksson, S., et al.\ 2003, \aap, 405, 15 

\noindent
Portegies 
Zwart, S.~F., Baumgardt, H., Hut, P., Makino, J., \& McMillan, S.~L.~W.\ 
2004, \nat, 428, 724 

\noindent
Reid, I.~N., Gizis, J.~E., 
\& Hawley, S.~L.\ 2002, AJ, 124, 2721 

\noindent
Rockenfeller, B., 
Bailer-Jones, C.~A.~L., \& Mundt, R.\ 2006, A\& A, 448, 1111 

\noindent
Sakamoto, T., Chiba, 
M., \& Beers, T.~C.\ 2003, A\&A, 397, 899 

\noindent
Saslaw, W.~C., 
Narasimha, D., \& Chitre, S.~M.\ 1985, \apj, 292, 348 

\noindent
Smith, M.~C., Mao, S., \& 
Wo{\' z}niak, P.\ 2003, \apjl, 585, L65 

\noindent
Stubbs, C.~W., Sweeney, 
D., Tyson, J.~A., \& LSST 2004, American Astronomical Society Meeting 
Abstracts, 205,

\noindent
Tonry, J. L., \& Schneider, D. P. 1988, AJ, 96, 807

\noindent
Tully, R.~B.\ 1988, Journal of 
the British Astronomical Association, 98, 316

\noindent
 Turner, E.~L., Wardle, 
M.~J., \& Schneider, D.~P.\ 1990, \aj, 100, 146 

\noindent
Udalski et al, 1994, Ap J 436, 103

\noindent
Uglesich, R.~R., 
Crotts, A.~P.~S., Baltz, E.~A., de Jong, J., Boyle, R.~P., \& Corbally, 
C.~J.\ 2004, \apj, 612, 877 

\noindent
Volonteri, M., Madau, 
P., \& Haardt, F.\ 2003, \apj, 593, 661 

\noindent
Walsh, D., 
Carswell, R.~F., \& Weymann, R.~J.\ 1979, \nat, 279, 381 

} 

\bigskip
\bigskip

\vspace{-0.1cm}
\noindent{Acknowledgements: 
I thank Ethan Vishniac,  for discussions which
encouraged me to split one long paper into this paper and its
companion paper. I would also like to thank Tim Brown, 
Lorne Nelson, and Eric Pfahl for 
discussions, and John Tonry for providing relevant information.  
This work was supported in part by NAG5-10705
and in part by the 
National Science Foundation under Grant No. PHY99-07949.        
} 
\end{document}